\documentclass[aps, twocolumn, showpacs, amsmath, amssymb, tightenlines, longbibliography, superscriptaddress]{revtex4-2} %,twocolumn, pre or preprint
\usepackage{empheq}
\usepackage{cases}
\usepackage{amsmath}
\usepackage{floatrow}
\usepackage{wasysym}
\usepackage{mathtools}
\usepackage[latin1]{inputenc}
\usepackage{graphicx} % Include figure files
\usepackage{dcolumn} % Align table columns on decimal point
\usepackage{bm} % bold math
\usepackage[dvipsnames]{xcolor}
\definecolor{darkblue}{rgb}{0,0,0.6}
\definecolor{darkred}{rgb}{0.6,0,0}
\usepackage[colorlinks=true,urlcolor=darkblue,citecolor=darkblue, linkcolor=darkred,hyperfootnotes=false]{hyperref}
\usepackage[textsize=small]{todonotes}

\usepackage{enumerate}
\usepackage{siunitx}
\usepackage{yhmath}

\newcommand {\ts}[1]{\textsubscript{#1}}

\newcommand{\rgeom}{r}

\newcommand{\cA}{\mathcal{A}}

\newcommand{\dd}{\text{d}}

\providecommand{\abs}[1]{\left| #1 \right|}

\newcommand{\ind}[1]{_{\mathrm{#1}}}

\newcommand{\ee}{\boldsymbol{e}}

\newcommand{\ff}{\boldsymbol{f}}

\renewcommand{\ggg}{\boldsymbol{g}}

\newcommand{\rr}{\boldsymbol{r}}

\newcommand{\pphi}{\boldsymbol{\phi}}

\newcommand{\ls}{l\textsubscript{s}}

\begin{document}

\title{
Interplay of gross and fine structures in strongly-curved sheets\\
}

\author{Mengfei He}
\email{mhe100@syr.edu}
\affiliation{Department of Physics, Syracuse University, Syracuse, NY 13244}
\affiliation{BioInspired Syracuse: Institute for Material and Living Systems, Syracuse University, Syracuse, NY 13244}

\author{Vincent D\'emery}
%\email{vincent.demery@espci.psl.eu}
\affiliation{Gulliver, CNRS, ESPCI Paris, PSL Research University, 10 rue Vauquelin, 75005 Paris, France}
\affiliation{Univ Lyon, ENS de Lyon, Univ Claude Bernard Lyon 1, CNRS, Laboratoire de Physique, F-69342 Lyon, France}
\author{Joseph D. Paulsen}
\affiliation{Department of Physics, Syracuse University, Syracuse, NY 13244}
\affiliation{BioInspired Syracuse: Institute for Material and Living Systems, Syracuse University, Syracuse, NY 13244}

%\date{\today}

\begin{abstract} %currently 249/250 words
Although thin films are typically manufactured in planar sheets or rolls, they are often forced into three-dimensional shapes, producing a plethora of structures across multiple length-scales. Existing theoretical approaches have made progress by separating the behaviors at different scales and limiting their scope to one. Under large confinement, a geometric model has been proposed to predict the gross shape of the sheet, which averages out the fine features. However, the actual meaning of the gross shape, and how it constrains the fine features, remains unclear. Here, we study a thin-membraned balloon as a prototypical system that involves a doubly curved gross shape with large amplitude undulations. By probing its profiles and cross sections, we discover that the geometric model captures the mean behavior of the film. We then propose a minimal model for the balloon cross sections, as independent elastic filaments subjected to an effective pinning potential around the mean shape. This approach allows us to combine the global and local features consistently. Despite the simplicity of our model, it reproduces a broad range of phenomena seen in the experiments, from how the morphology changes with pressure to the detailed shape of the wrinkles and folds. Our results establish a new route to understanding finite buckled structures over an enclosed surface, which could aid the design of inflatable structures, or provide insight into biological patterns. 
\end{abstract}

\maketitle

Complex patterns of wrinkles, crumples and folds can arise when a thin solid film is stretched~\cite{Cerda02, Cerda03, Davidovitch11}, compacted~\cite{Matan02, Blair05, Vliegenthart06, Witten07}, stamped~\cite{Hure12, Roman12, Tobasco22}, or twisted~\cite{Chopin13, Chopin15, Legrain16}. %delamination~\cite{Vella09a, Hure11}, 
These microstructures arise to solve a geometric problem: they take up excess length at a small scale to facilitate changes in length imposed at a larger scale, by boundary conditions at the edges or by an imposed metric in the bulk. 
When the confining potential is sufficiently soft, like that presented by a liquid, the sheet can have significant freedom to select the overall response that the small scale features decorate \cite{Holmes10, Chopin13, Vella15, Paulsen15, Paulsen19, Siefert19, Ripp20, Timounay21}. 
Understanding how gross and fine structures are linked, especially in situations with large curvatures and compression, remains a frontier in the mechanics and geometry of thin films. 

To date, the dominant approach has been to treat gross and fine structure separately. 
For example, tension-field theory~\cite{Mansfield89} accounts for the mechanical effect of wrinkling on the stress and strain fields, while ignoring the detailed deformations at the scale of individual buckles. 
Recent work~\cite{Davidovitch11} has established how to calculate the energetically-favored wrinkle wavelength anywhere within a buckled region. 
But owing to geometric nonlinearities, these approaches have been limited to situations with small slopes and a high degree of symmetry, and they assume small-amplitude sinusoidal wrinkles at the outset. 
A geometric model was recently developed~\cite{Gorkavyy10, Paulsen15} for situations where an energy external to the sheet - like a liquid surface tension or gravity - ultimately selects the gross shape. 
Although this model can address situations with arbitrary slopes, it is typically agnostic to the exact form of the fine structures. Moreover, it is unclear what the gross shape precisely corresponds to.

At the small scale, significant progress has been made on understanding oscillating buckled features by analyzing an inextensible rod attached to a fluid or solid foundation \cite{Cerda03, Paulsen16}. 
This approach can predict the energetically-favored wavelength of monochromatic wrinkles, but much less is known about the selection of more complex or evolving microstructures. 
Morphological transitions, like the formation of localized folds, have been largely analyzed on planar substrates \cite{Diamant11, Brau11, Demery14, Oshri15}, and it is not known how they are modified when the gross shape is curved and can freely deform \cite{Holmes10}. 

\begin{figure}[tb]
\includegraphics[width=0.98\textwidth]{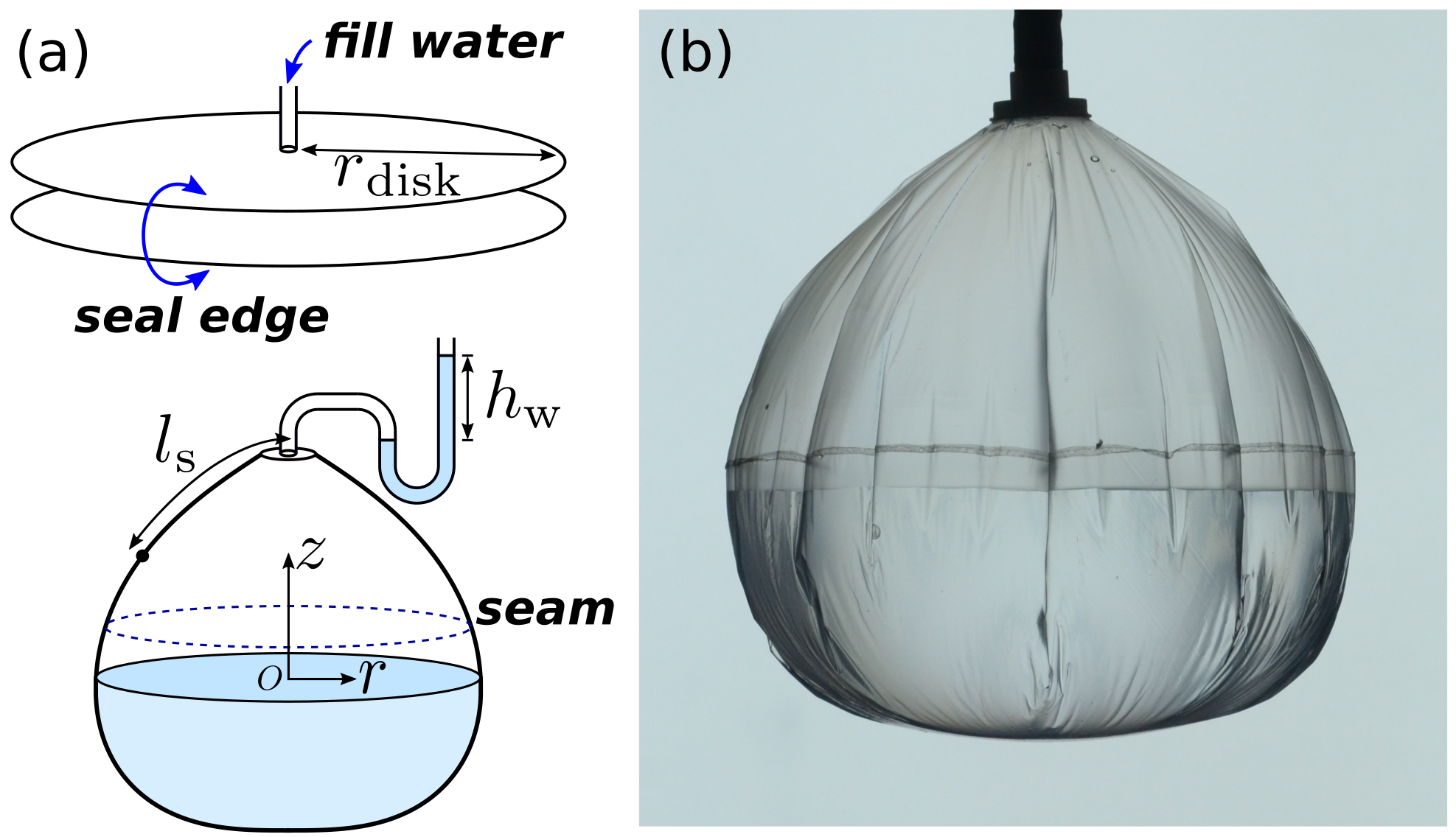} 
\caption{(a) We make partially-filled water balloons by sealing two circular sheets of radius $r_\text{disk}$ that bend easily but strongly resist stretching. We vary the air pressure inside the balloon, which we measure using the height $h_\text{w}$ of a water column in a U-shaped tube. 
We use cylindrical coordinates $(r,z)$, and we denote the arclength from the top of the balloon as $l_\text{s}$. 
(b) Side-view photograph of the setup with water volume $V\ts{water}=277.9$ mL and air pressure $p=276$ Pa.}
\label{fig:schematic}
\end{figure}

Here we elucidate the interplay of the gross and fine structures of a strongly deformed thin sheet by studying a water balloon made by unstretchable thin membranes (Fig.~\ref{fig:schematic}). 
We obtain a wide range of deformations by varying the internal air pressure and the volume of liquid in the balloon. 
Despite the complex surface arrangements, a simple geometric model omitting the bending cost captures accurately the mean, or the \textit{gross shape} of the balloon membrane measured from extracting the cross sections.  
The observed side \textit{profile} of a balloon, on the other hand, may differ from the mean shape due to surface fluctuations. 
We show how the finite wrinkle amplitude can be accounted for and combined with the mean behavior to predict the outer envelope of the wrinkled balloon, in agreement with our experiments. 
To understand the buckled microstructure in more detail, we model a cross section of the balloon as an effective filament pinned to the prediction of the geometric model.
Remarkably, our parsimonious model quantitatively captures various observed surface morphologies of the balloon, while retaining the agreement between the mean shape and the prediction of the geometric model. 
We measure the size of the self-contacting loops at the tips of the folds that form at high pressures, which further corroborates the two-dimensional behavior of the cross sections as elastic filaments. 
These results provide a paradigmatic example of how to analyze gross shape and fine structures in a unified approach, for strongly-curved sheets.

\section{Experimental setup}

We construct closed membranes by sealing together disks of initially-planar plastic sheets.  The disks are cut from polyethylene produce bags of thickness $t = 8.0 \pm 0.8$ $\mu$m and Young's moduli varying from $E=315$ MPa to $E=1103$ MPa as a function of the angle, or from plastic food wraps with $t=10.2 \pm 0.8$ $\mu$m and Young's moduli varying from $E=133$ MPa to $E=195$ MPa as a function of the angle. The measurements are detailed in SI.
An air-tight seal is formed by pressing a heated iron ring of diameter $166$ mm.  
The fused circular double layers are then cut out, and a nylon washer of radius $r\ts{top}$ is glued to the center of the top layer for hanging the bag and connecting a tube that can supply air and water into the bag. 
The same tube is connected to a custom manometer made from two plastic cylinders connected by a U shaped tube; the difference in water height $h\ts{w}$ allows us to measure the overpressure via the hydrostatic pressure difference $\rho g h\ts{w}$ [Fig.~\ref{fig:schematic}(a)]. 

Once inflated, the balloon transforms from a flat initial state to a strongly curved global shape with a complex arrangement of surface structures [Fig.~\ref{fig:schematic}(b)]. 
Side views are photographed with a Nikon DLSR with back lighting from an LED white screen. 
Despite the anisotropy of the Young's modulus, we do not see large systematic variations in the morphological behaviors as a function of angle of loading. 

We also measure horizontal cross sections of the balloons by scattering a laser light sheet into the system [Fig.~\ref{fig:fine}a]. 
We photograph the scattered laser light from the side at an oblique angle. 
A calibrated perspective transformation is then applied to produce the final horizontal image. 
To better capture the back of the balloon, we reflect a portion of the light sheet onto the back of the system with two vertical mirrors. 
The above procedure is sometimes repeated from different azimuthal angles and the results are superimposed, to reduce noise and better capture the back of the balloon. 
The images captured from different angles collapse well on one another, indicating that the cross sections are obtained accurately without geometric distortion.

\begin{figure*}[tb]
	\includegraphics[width=0.98\textwidth]{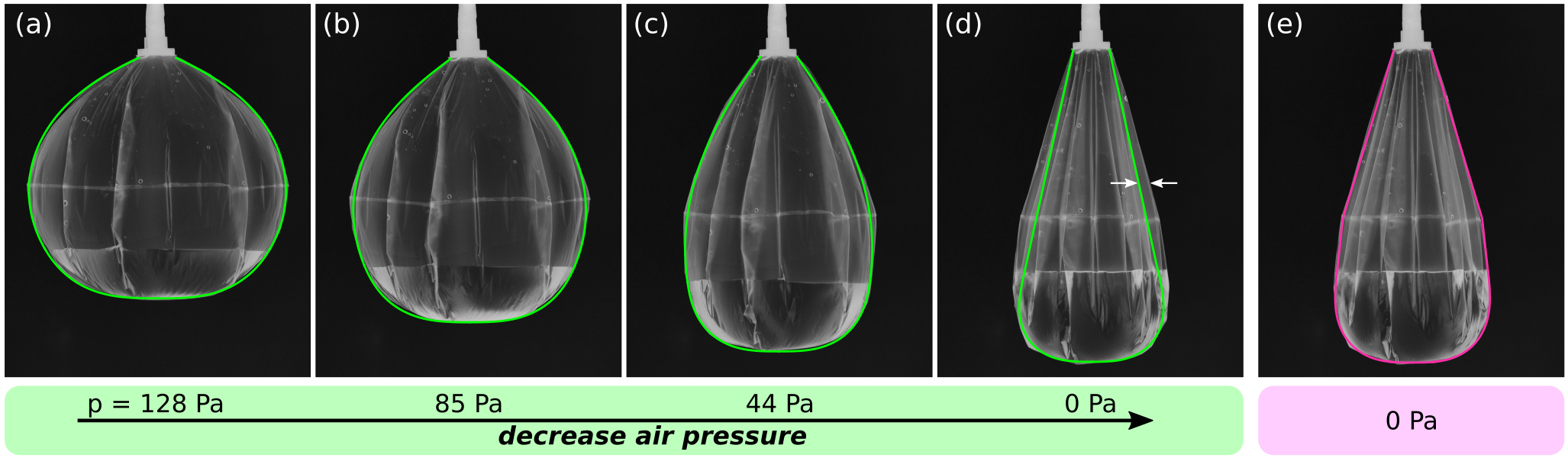}
	\caption{(a-d) Side-view photographs of a balloon filled with $V\ts{water}=100.4$ mL of water at various internal pressures from $p=128$ Pa to 0 Pa. Image grayscale inverted for clarity. 
	Green curves: Predictions from the geometric model (Eqs.~\ref{eq:geom_model}) at the corresponding pressures with no free parameters. 
	The agreement is very good at high pressures, but there is a significant discrepancy at $p=0$. 
	(e) Pink curve: Adding the amplitude of the wrinkled envelope (Eq.~\ref{eq:linear}) to the geometric model gives good agreement with the side-view profile of panel (d). 
	}
	\label{fig:gross}
\end{figure*}

%The profile of a balloon compared with the prediction given by the geometric model, showing a good match at high pressures and a mismatch at low pressures.

\section{Geometric model for gross shape}

To capture the side-view profiles of the balloons, we use a simple \textit{geometric model}~\cite{Gorkavyy10, Pak10, Paulsen15} that idealizes the balloon as a smooth, axisymmetric surface $\rgeom(z)$ with no surface fluctuations [Fig.~\ref{fig:schematic}(a)]. 
This effective surface is free to bend and cannot carry compressional stresses as they are relieved by the wrinkles and folds around the balloon. 
Assuming the entire balloon is wrinkled so that the circumferential tension vanishes everywhere, the only remaining tension is the effective longitudinal stress $T\ts{eff}=T\ts{s}l\ts{s}/\rgeom$, where $T\ts{s}$ is the physical longitudinal stress in the balloon membrane, and $l\ts{s}$ the meridian arc-length from the top center to the point in question [Fig.~\ref{fig:schematic}(a)]. 
Force balance~\cite{Taylor63, Smalley64, Mansfield89, Baginski98, Baginski04, King12} for the effective surface gives:
\begin{align}
	\label{eq:normal}
	\kappa T\ts{eff} &= p\\
	\label{eq:horizontal}
	\partial_{\rgeom} (\rgeom T\ts{eff}) &= 0,
\end{align}
where $\kappa = -\rgeom''/(1+\rgeom'^2)^{3/2}$ is the curvature of the profile, 
%\note{if helpful, consider replacing with $\rgeom$ for easy notation changes down the road}, 
and  $p=\rho gh\ts{w}$ or $\rho g(h\ts{w}-z)$ is the local pressure drop across the membrane, above or below water.
We set $z=0$ as the water level. 

Equation~\ref{eq:normal} is the normal force balance, involving only the curvature of the profile due to the absence of azimuthal tension.  
Equation~\ref{eq:horizontal} is the horizontal force balance, which immediately leads to
\begin{align}
	T\ts{eff} = \frac{f}{\rgeom},
	\label{eq:t}
\end{align}
where $f$ has the dimension of force.  
%giving rise to a length scale $l_0 = \sqrt[3]{f/(\rho g)}$.  
Substituting Eq.~\ref{eq:t} into Eq.~\ref{eq:normal} gives
\begin{subequations}\label{eq:geom_model}
    \begin{empheq}[left={\begin{aligned}-\frac{f}{\rho g}\frac{\rgeom''}{\rgeom(1+\rgeom'^2)^{3/2}} =\end{aligned} \empheqlbrace}]{align}
        &h\ts{w} &0<z<z\ts{top}& \label{eq:above}
        \\
        &h\ts{w}-z &z\ts{bot}<z<0&. \label{eq:below}
    \end{empheq}
\end{subequations}  
We denote $z\ts{top}$, $z\ts{bot}$ as the top and bottom coordinates of the balloon.  Equations~\ref{eq:geom_model} are then two second-order ODE's with three parameters $f$, $z\ts{top}$ and $z\ts{bot}$ unknown \textit{a priori}, and hence should be supplemented with seven boundary conditions: $\rgeom(0_+)=\rgeom(0_-)$ and $\rgeom'(0_+)=\rgeom'(0_-)$ at the water level, $\rgeom(z\ts{top})=r\ts{top}$, $\rgeom(z\ts{bot})=0$ and $\rgeom'(z\ts{bot}) \rightarrow \infty$ at the top and the bottom of the balloon, together with the sheet inextensibility constraint $r\ts{top}+\int_{z\ts{bot}}^{z\ts{top}}{\sqrt{1+\rgeom'^2}}\mathrm{d}z = 2r\ts{disk}$ and a prescribed water volume $\int_{z\ts{bot}}^0{\pi \rgeom^2}\mathrm{d}z=V\ts{water}$.

Equations~\ref{eq:geom_model} are integrated numerically using \textit{odeint} implemented in the \textsc{SciPy} package \textit{integrate}, and the solutions, which we denote as $r\ts{gm}(z)$, are plotted in Fig.~\ref{fig:gross} over the corresponding images. 
This comparison is done with no free parameters. 
%to produce each numerical curve we require as inputs the disk radius, the radius of the washer at the top of the balloon, and the volume of water in the bag, measured on a scale before filling into the balloon. 
There is an excellent agreement between the numerical predictions and the apparent shape of the balloon at high pressure, despite the complex surface arrangements. 
However, the agreement is poor at low pressure, as shown by the $p=0$ case in Fig.~\ref{fig:gross}(d).  To understand this apparent discrepancy at low air pressures we turn to the investigation of the fine structure of the balloon.

\begin{figure*}[tb]
	\includegraphics[width=1\textwidth]{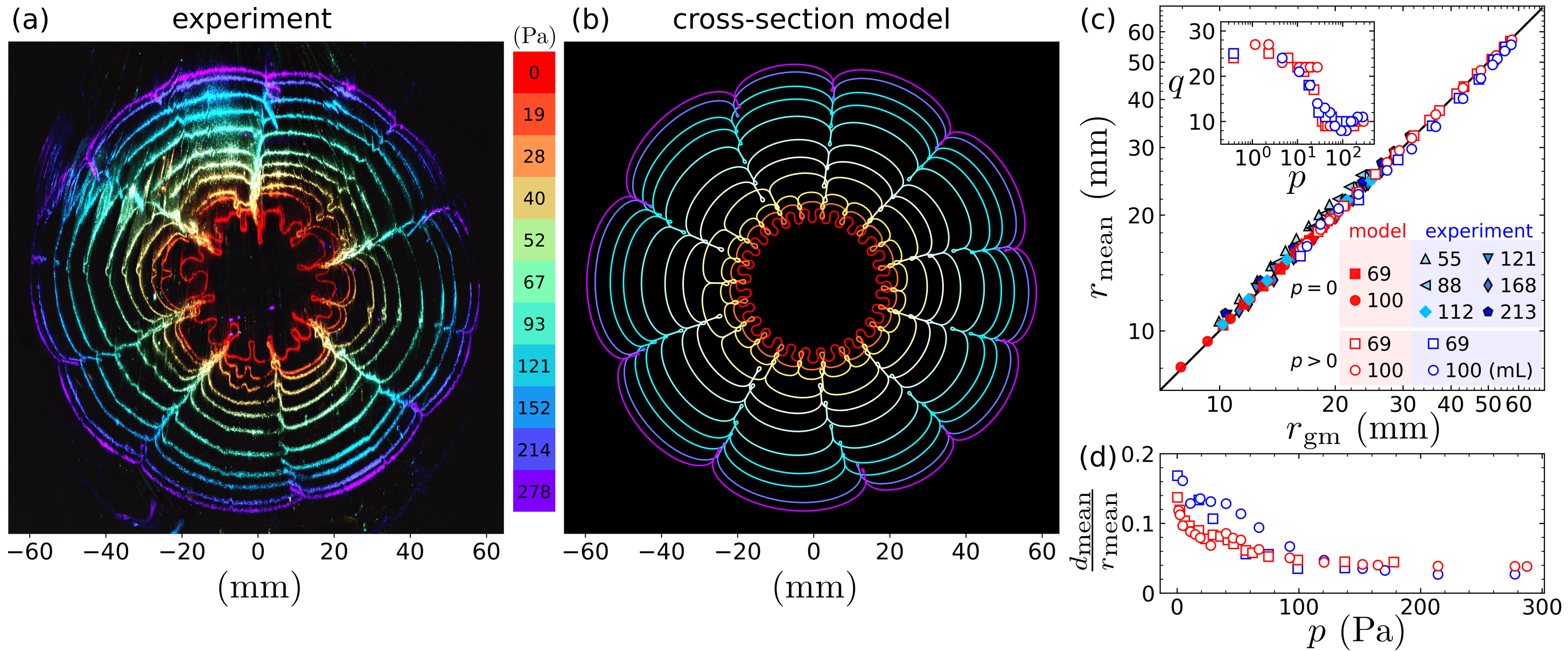}
	\caption{
		Cross-section model captures the morphological features of the balloon cross sections.
		(a) Cross-section shapes measured by scattering a sheet of laser at a fixed distance $z\ts{top}-z=39.4$ mm below the top of a balloon with $V\ts{water}=100.0$ mL, which is gradually deflating from $p=278$ Pa to 0 Pa. The photographs are tinted and superimposed to compare the different morphologies.
		(b) Shapes from the cross-section model. All physical parameters (pressure, bending modulus, length of cross sections, volume of water) were set to match the conditions in panel (a), and a single $\eta=15$ in Eq.~\ref{eq:U_t} was selected by matching the wavenumber with experiment at high pressures. 
		(c) The average distance of a cross section to its centroid, $r\ts{mean}$, versus the prediction of the geometric model, $r\ts{gm}$, in the same $z-$plane. The data span various cross sections in different balloons made by produce bags (markers with black edges) or food wrap. Blue markers: experimental measurements. Red markers: cross-section model.  Filled markers: $p=0$.  Open markers: $p>0$.
		Inset: wavenumber $q$ versus pressure, showing that the cross-section model captures the wrinkle-fold transition. 
		(d) The average fractional protrusion of a cross section, $d\ts{mean}/r\ts{mean}$, versus pressure. 
		The cross-section model reproduces the trend in the experiments, where the data decay quickly upon inflation. 
		This trend explains how the geometric model can match the side-view profiles at high pressures [Fig.~\ref{fig:gross}(a-c)]. 
		%The cross-section model captures the quick decay of $d\ts{mean}$ upon inflation. 
		%\note{needs caption -- just what is being plotted, any technical details that don't deserve main-text treatment}	
	}
	\label{fig:fine}
\end{figure*}

\section{Cross sections}
Cross sections at a fixed vertical distance from the top are colorized, and superimposed to highlight the morphological change as a function of pressure. [Fig.~\ref{fig:fine}(a)]. 
The image shows a transition in the fine structure, starting from folds at large pressure to wrinkles at lower pressure, with an increase of the wavenumber. 

For each contour, we identify the center of the balloon with the centroid of the contour, and then measure $r\ts{mean}$: the mean distance between the contour and the center of the balloon.  Figure~\ref{fig:fine}(c) compares the measured $r\ts{mean}$ versus the predicted radius of the gross shape $r\ts{gm}$ in the same plane. 
The agreement shows that the geometric model accurately predicts the \textit{mean} shape of the balloon surface. 
This agreement is robust; the data include cross sections at multiple vertical locations in the deflated balloon, and different pressures in the inflated balloons, and we have also varied the membrane material and water volume. 
%for $p=0$, we collect many data points for each static balloon shape, by taking cross sections at multiple heights; 
%for $p>0$, we collect data points at multiple pressures. 

To quantify the difference between the mean shape of the balloon and the apparent profile, we measure the average protrusion $d\ts{mean}$ of each cross section, which we define to be the average amplitude of the local maxima of each contour.  Figure~\ref{fig:fine}(d) shows that $d\ts{mean}$ decays rapidly with increasing pressure. 
Taken together, the results in Figs.~\ref{fig:fine}(c,d) resolve the apparent discrepency at low pressure between the geometric model and the side-view profiles in Fig.~\ref{fig:gross}. 
Namely, the side-view profile is given by the mean shape plus the amplitude of the wrinkly undulations, $r\ts{gm}+d\ts{mean}$. 
Figure~\ref{fig:fine}(d) shows that these undulations can be rather large at zero pressure --- more than 10\% of the mean --- but they become much smaller when there is an internal pressure. 
%Note that the side-view images show the apparent shape $\approx r\ts{gm}+d\ts{mean}$.  
%The decay in $d\ts{mean}$ quantitatively explains that the prediction of the geometric model matches the apparent shape at large pressures, while there is a significant difference at low pressures.  
To quantify the deviation $d\ts{mean}$, and understand how it arises from the particular curve shapes, we now study the $p=0$ case in more detail.

\begin{figure*}[tb]
	\includegraphics[width=0.95\textwidth]{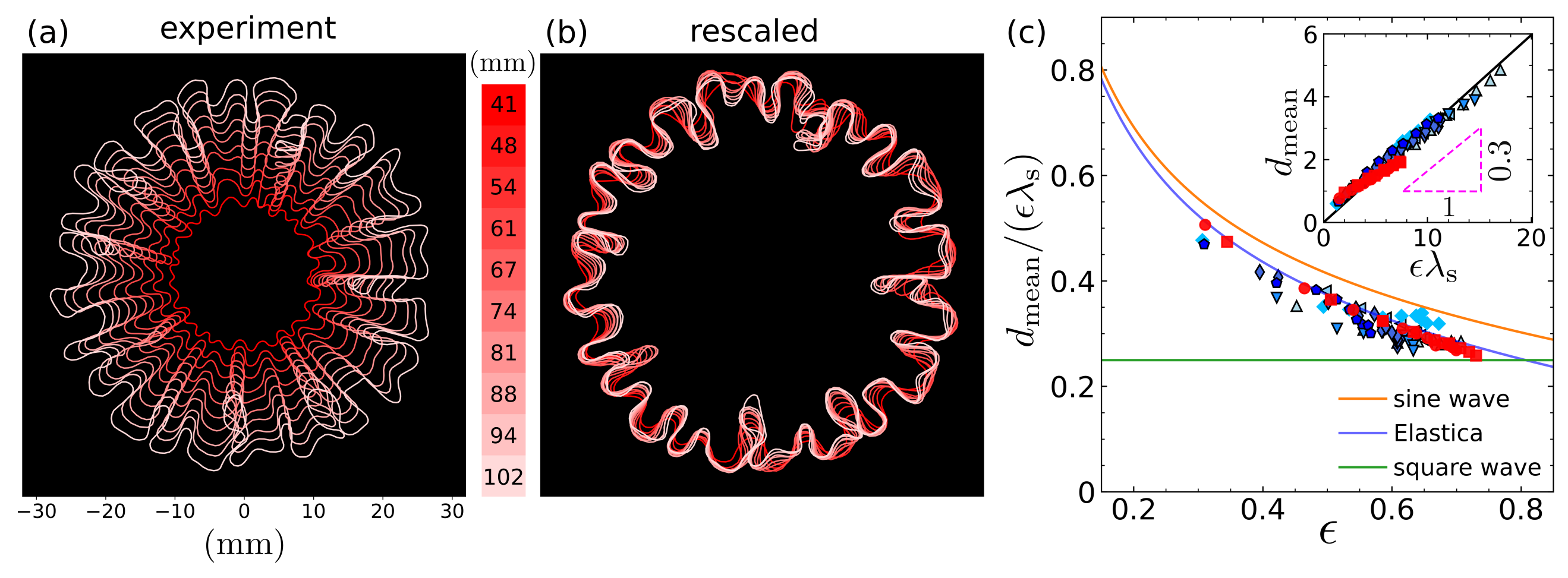}
	\caption{
	Cross sections in a deflated balloon exhibits a generalized conical structure.  
	(a) Cross-section curves extracted from an balloon with  $V\ts{water}=112.9$ mL at $p=0$, at 10 different heights. 
	The contour is traced and plotted as a line for clarity. 
	%The apex of the conical side profile of the balloon is extracted and the cross sections are colored by shades of red representing its vertical distance $h$ to the apex. 
	(b) Cross sections rescaled by the distance to the extrapolated apex of the conical profile (this distance is denoted in the colorbar). 
	This operation nearly collapses the rescaled shapes, indicating a generalized cone structure of the balloon structure. 
	(c) Inset: the average protrusion, $d\ts{mean}$, of the cross sections versus $\epsilon \lambda\ts{s}$ from experiments (blue) and the cross-section model (red). See Fig.~\ref{fig:fine}(c) for full legend. 
	The trend is approximately linear. 
	Main: to examine these data in more detail, we plot $d\ts{mean}/(\epsilon \lambda\ts{s})$ versus the effective strain $\epsilon$, for various balloons at $p=0$ from both the experimental measurements and the numerical predictions of the cross-section model. 
	For comparison, we plot the exact relations for a sine wave, square wave, and Euler's \textit{Elastica}, where in each case the amplitude is controlled by $\epsilon$. 
	The \textit{Elastica} shows the best agreement with the data. 
	}
	\label{fig:zero_p}
\end{figure*}

\section{Wrinkle shape at zero pressure}

Cross sections of a deflated water balloon measured at equal vertical intervals above the seam are shown in Fig.~\ref{fig:zero_p}(a). 
Noting that the apparent profile of the top half of the balloon looks to be conical, we extrapolate an apex of the apparent profile from the side-view image. 
We then rescale the cross sections by the distance to this apex. 
This simple rescaling nearly collapses all the cross sections [Fig.~\ref{fig:zero_p}(b)], indicating that not only the gross shape but also the fine structure has the shape of a generalized cone. 

A consequence of a conical structure is that the wrinkle wavenumber does not vary significantly with height $z$. 
To understand this observation, we first assume that the selection of the wavelength is local \cite{Paulsen16} and dominated by the tensional substrate stiffness \cite{Cerda03}. 
Then the wavelength would scale as: 
%The scaling of the wavelength reads~\cite{Cerda03}: 
$\lambda\sim(B/T\ts{s})^{1/4}r\ts{disk}^{1/2}$. %, assuming the surface undulations emerge as the \textit{tension wrinkles}, namely, as a result of the balance between the bending cost of the membrane and the meridian tension. 
On the other hand, $T\ts{s}$ varies with $z$, and so should $\lambda$, leading to a mismatching wavelength along the balloon meridian. 
The resulting length scale $\ell$ associated with a wavenumber change scales as $\ell \sim \lambda^2\sqrt{T\ts{s}/B}$~\cite{Vandeparre11}. 
Taken together, $\ell \sim r\ts{disk}$, which contradicts our assumption that the selection of wavelength is local, and explains the invariance of the wavenumber within the size of the balloon. 
%Note that taking $\lambda\sim(B/T\ts{s})^{1/4} l_\text{s}^{1/2}$ in the argument above (where $r\ts{disk}$ is replaced with $l_\text{s}$) leads to yet another dependence on $z$, which we do not see. 

To study the shape of the individual cross sections, we investigate the relations among the observables of the mean maximum amplitude $d\ts{mean}$, the wrinkle wavelength and how much the cross section is compressed. 
We define the \textit{material wavelength} $\lambda\ts{s}\equiv2\pi l\ts{s}/q$, which divides the total arclength of the cross section by the number of undulations $q$ measured by counting peaks. 
We quantify the amount of compression by the effective strain $\epsilon\equiv(l\ts{s}-r\ts{mean})/l\ts{s}$ (to be contrasted with the local material strain, which is vanishingly small for our buckled films). 
Note that $\lambda\ts{s}$ approaches the ``usual wavelength", $\lambda$, when $\epsilon$ approaches $0$. 
The particular shape of the undulations tells how $d\ts{mean}$, $\lambda\ts{s}$, and $\epsilon$ are related. 
For example, small amplitude sinusoidal waves obey $d\ts{mean}\propto\sqrt{\epsilon}\lambda\ts{s\\}$ and for square waves $d\ts{mean}=\epsilon\lambda\ts{s}/4$. 
To examine the relation in our system, we plot $d\ts{mean}$ versus $\epsilon\lambda\ts{s}$ in the inset to Fig.~\ref{fig:zero_p}(c). 
Remarkably, the data collapse onto a line that is fit well by:
\begin{align}
	d\ts{mean} \approx \beta \epsilon\lambda\ts{s},
	\label{eq:linear}
\end{align}
where $\beta$ is the linear coefficient with a best-fit value of $0.30$. %$\beta\ts{fit}\approx0.298$

The simple relation of Eq.~\ref{eq:linear} %immediately implies an empirical estimation of the linear correction, 
shows how to readily estimate the amplitude of wrinkles $d\ts{mean}$ as a function of $z$. 
Given a volume of water, a washer radius, and a bag size, one may compute the arclength $l_s(z)$ and the mean radius $r\ts{mean}\approx r\ts{gm}(z)$ for the gross shape. 
The crucial ingredient from the experiment is the observed wavenumber $q$ (which is approximately independent of $z$). 
With just these parameters, one may then obtain $\epsilon(z)$ and $\lambda\ts{s}(z)$, which combine via Eq.~\ref{eq:linear} to give $d\ts{mean}(z)$. 
This wrinkled envelope can be added to the gross shape from the geometric model to yield a predicted apparent shape. 
We do this in Fig.~\ref{fig:gross}(e); the result matches the experimental profile very well, especially when compared to the geometric model without the wrinkled envelope for the same balloon, in Fig.~\ref{fig:gross}(d). 
%$d\ts{mean}$ for any balloon, by computing $\epsilon$ and $\lambda\ts{s}$ from a side-view image alone. 
%As an example, we show in Fig.~\ref{fig:gross}(e) that Eq.~\ref{eq:linear} indeed accounts for the difference between the apparent shape and the prediction of the mean shape given by the geometric model of Fig.~\ref{fig:gross}(d). 
The correction continues to be favorable past the seam through the bottom of the balloon, which offers an explanation for the kink in the  apparent profile at the location of the seam where the two disks are heat-sealed together. 
Namely, this kink can be understood as arising from the triangular peak in the amount of material to be packed into the confined gross shape, since the excess length as a function of $z$ grows linearly up to the location seam, and then falls linearly past the seam. 
Thus, it is a geometric and not a mechanical effect, that one might expect could arise due to the relatively larger rigitidy of the heat-sealed seam (see SI). 
%is not caused by the seam rigidity (more details in SI), but by a kink in the amount of available material to be packed into the confined gross shape $r\ts{mean}$. 

To examine the curves in more detail, we plot the ratio $d\ts{mean} / \epsilon\lambda\ts{s}$ as a function of $\epsilon$ in Fig.~\ref{fig:zero_p}(c). 
For comparison, we show the curves for sinusoidal waves, square waves, and the \textit{inflexional Elastica}~\cite{Love1906}. 
The data are close to the trend for the Elastica -- local minimizers of the integrated squared curvature, which capture the bending mechanics of an inextensible rod in equilibrium. 
This behavior is expected for confined isometric surfaces~\cite{Cerda05}; in our case, the presence of tension along the wrinkles may introduce deviation to pure Elastica, as we will see. 

Our detailed analyses of the cross sections have two main conclusions: 
First, in contrast to hierarchical wrinkling such as in a light suspended curtain~\cite{Schroll11, Vandeparre11}, here the cross-section shape is approximately independent of $z$ [Fig.~\ref{fig:zero_p}(b)]. %, suggesting that that the cross-sectional shape is determined primarily by mechanics \textit{within} a cross-section. 
Second, the shared properties of the cross sections with \textit{Elastica} points to the relevance of a minimization of the bending energy \textit{within} a cross section. 
%These findings motivate a quasi-two-dimensional model for the cross section. 
These findings motivate a quasi-two-dimensional model for a typical cross section, which is representative of all the others in a given configuration of the balloon.

\section{Quasi-two-dimensional model}

We model a cross section of the balloon at $l\ts{s}$ as an inextensible filament of total length $L=2\pi l\ts{s}$.  For a filament with a two-dimensional, arc-length parametrized configuration $\bm{r}(s)$, the bending energy is 
\begin{align}
U\ts{b}=\frac{B}{2}\int\bm{r}''(s)^2\mathrm{d}s,
\label{eq:U_b}
\end{align}
where $B$ is the bending rigidity. 

Inspired by the ``elastic foundation'' effect induced by a longitudinal tension~\cite{Cerda03}, we introduce a confinement energy
\begin{align}
U\ts{t}=\frac{\eta T_s}{2r\ts{disk}^2}\int[\bm{r}(s)-\bm{r}\ts{id}(s)]^2\mathrm{d}s,
\label{eq:U_t}
\end{align}
penalizing deviations of the filament from an ideal configuration $\bm{r}\ts{id}(s)=r\ts{conf}\cdot(\cos(s/L), \sin(s/L))$.  Here $r\ts{conf}$ is a parameter to be determined by the solution of the geometric model for the gross shape of the balloon.  Crucially, we further incorporate the finite slope effect by penalizing both the radial and azimuthal deviations in the integrand of Eq.~\ref{eq:U_t}, without which the system configuration $\bm{r}(s)$ would have favored a single deep fold~\cite{Pocivavsek08, Demery14}.  To account for the double-curved profile and the boundary connecting to the balloon membrane below the water level, we introduce a dimensionless factor $\eta$.  To retain the predictive power, we set $\eta$ to be a fixed numerical parameter independent of $p$, to be determined \textit{a posteriori} by matching the configuration wavenumber $q$ produced by the model with the experimental observation.  

So far the model is for $p=0$, but pressure can be incorporated by considering the work that it does on the filament. 
This work is $p$ multiplied by the enclosed area, which reads: 
\begin{align}
U\ts{p} = -\frac{p}{2}\int\bm{r}(s)\times\bm{r}'(s)\mathrm{d}s.
\label{eq:U_p}
\end{align}
The total energy associated with the model cross section is then
\begin{align}
	U\ts{tot}[\bm{r}(s)]= U\ts{b}+U\ts{p}+U\ts{t}.
	\label{eq:U_tot}
\end{align}

In computing $U\ts{t}$, the parameter $r\ts{conf}$ is specified by requiring the case of an inextensible balloon membrane with zero bending rigidity to recover the prediction of the geometric model.  Accordingly, for the representative filament with bending rigidity $B=0$, $\bm{r}(s)$ should approach a circle of radius $r\ts{gm}$ with arbitrarily small and dense oscillations.  The total energy becomes $\lim_{B\rightarrow0} U\ts{tot}(r) =(\eta T_s)/(2r\ts{disk}^2)(r-r\ts{conf})^2L-\pi p r^2$ and should take its minimum at $r=r\ts{gm}$.  Setting $\partial_r \lim_{B\rightarrow0} U\ts{tot}|_{r\ts{gm}}=0$ gives 
\begin{align}
	r\ts{conf} = r\ts{gm}\bigg(1 - \frac{1}{\eta}\frac{p r\ts{disk}^2}{T_s l\ts{s}}\bigg).
	\label{eq:rconf}
\end{align}
In this flexible membrane limit, the balloon assumes a radius of $r\ts{conf}$ at $p=0$, while for $p>0$ the pressure term $U\ts{p}$ displaces the balloon radius to $r\ts{gm}$.

We are interested in finding the configuration $\bm{r}(s)$ that minimizes the total energy $\delta U\ts{tot}[\bm{r}(s)]=0$.  To realize the energy minimization, we derive the corresponding force $\bm{f}\ts{tot}(s)=-\delta U\ts{tot}/\delta\bm{r}(s)$ and evolve the configuration using the dynamics of an over-damped system $\partial_t \bm{r}(s)=\bm{f}\ts{tot}(s)$ to equilibrium.  An ``inextensibility force'' is added to preserve the filament length $L$~\cite{Tornberg04}, as is a self-repulsion force against self-crossing (SI).

Figure~\ref{fig:fine}(b) shows the predictions of the filament model at various pressures.  Qualitatively, the numerical configurations capture the lobe formation at high pressure, the rounded wrinkle shape at low pressure, and a wrinkle-fold transition with wavenumber proliferation at intermediate pressure.  For each configuration, the mean radius $r\ts{mean}$, similarly defined as in the experimental measurements, is computed and plotted in Fig.~\ref{fig:fine}(c) as the red markers.  The numerical minimization reproduces quantitatively that the mean radius of the filament is close to the prediction of the geometric model.  
%Note that is a natural consequence of Eq.~\ref{eq:rconf}: the confinement restores $\bm{r}(s)$ \textit{toward} a smaller radius than $r\ts{gm}$, countering the expanding effect of the internal pressure, such that the filament ends up oscillating around $r\ts{gm}$.  
Similarly, the mean amplitude $d\ts{mean}$ of the filament protrusions is extracted and a quick decay over increasing pressure is manifest as shown in Fig.~\ref{fig:fine}(d).  
For all pressures, the wavenumber produced by the filament model increases with an increasing numerical coefficient $\eta$ in Eq.~\ref{eq:U_t}.  Nonetheless, the threshold pressure for the wrinkle-fold transition is insensitive to our choice of $\eta$ (see SI).  Such overall vertical shift of the numerical $q(p)$ curve enables the specification of $\eta$ by matching the corresponding curve obtained in experiments, shown in the inset of Fig.~\ref{fig:fine}(c). 

For the special case of $p=0$, we extract $d\ts{mean} / \epsilon\lambda\ts{s}$ and $\epsilon$ from the model filaments.  Figure~\ref{fig:zero_p}(d) shows that while the numerical model gives a relation quantitatively consistent with that of the experimental measurement, it further reveals the small but noticible deviation of the cross-section curves from pure \textit{Elastica}.  Such distinction is irresolvable experimentally due to system noise, yet shall be expected from the longitudinal tension along the conical shape.

\section{Shape of deep folds}

\begin{figure}[tb]
	\includegraphics[width=0.99\textwidth]{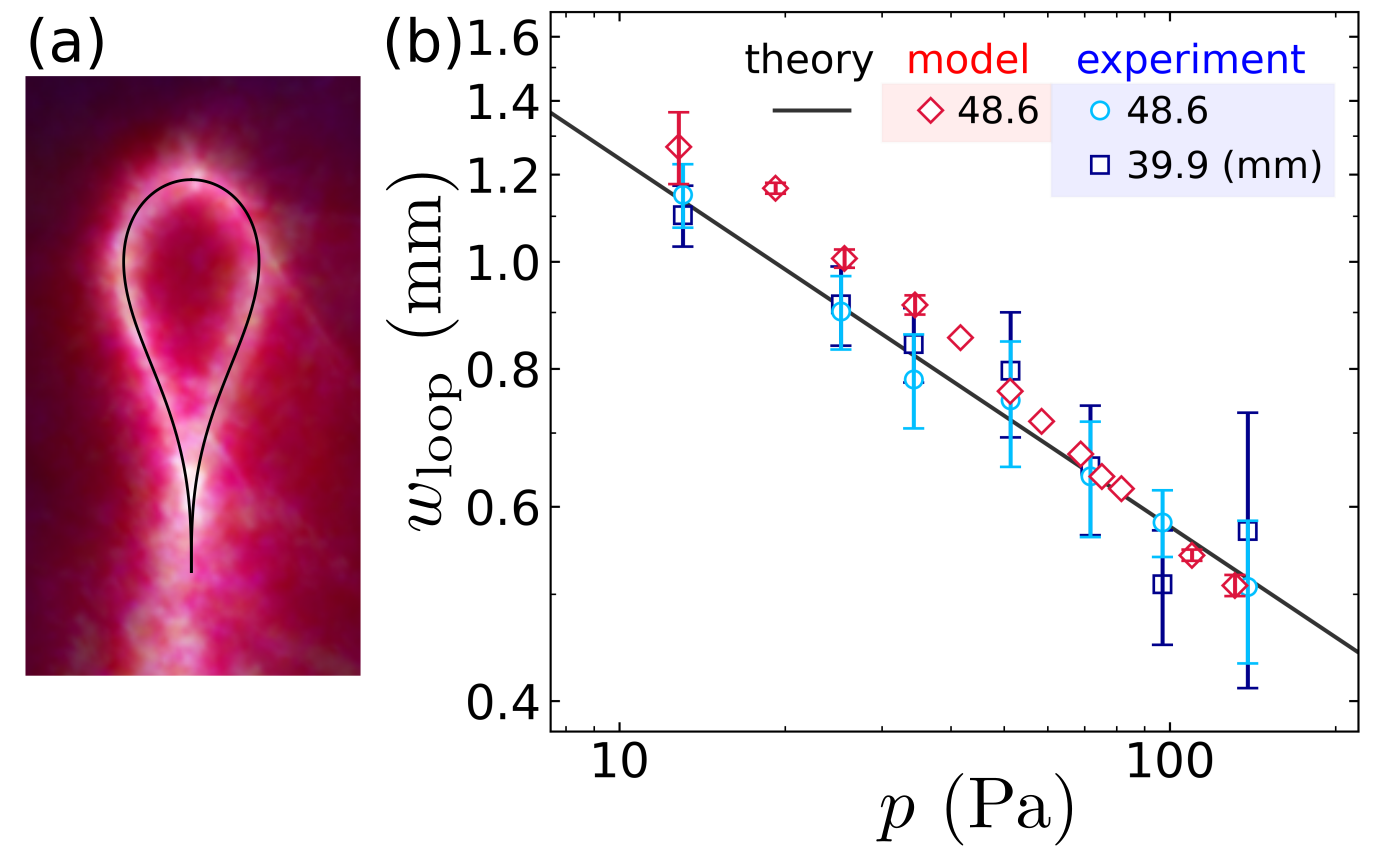}
	\caption{
		Loops at the tips of folds at high pressure. 
		(a) Superposition of 25 scaled laser-scattering images of fold tips, with $V\ts{water}=103.5$ mL and various pressures. 
		Black curve: exact similarity solution in the high pressure limit. 
		(b) The loop width, $w\ts{loop}$, versus $p$. 
		The data are an average of multiple loops at a given pressure; error bars show the variation in the observed loop size, quantified by the standard deviation over the mean of the measurements. 
		Experimental measurements at two different heights give consistent results. 
		Results from the cross-section model (red markers) in the same conditions agree well with the experiments. 
		Line: Exact model prediction in the high pressure limit, using the value of $B=1.58\times10^{-8}$ J that we measured with a tensile tester.
	}
	\label{fig:loops}
\end{figure}

We have seen that the minimal physical ingredients to reproduce the cross-section morphology are the effects of pressure, the bending energy within a cross section, and an effective pinning potential that captures the coupling of cross sections to the prediction of the geometric model. 
To investigate the balloon morphology further, we turn to study the shape of individual folds at large pressure, by considering these basic physical ingredients. 
At the tip of each fold, the membrane curves around sharply, shown in the cross section as a small loop. 
We observe that the size and shape of the folds can vary between different folds at different pressures. 
To see whether there is a commonality among the folds, we scale and superimpose the loops onto each other. 
We then add the brightness values of the photos to create a single composite image. 
Remarkably, a master shape appears, as shown in Fig.~\ref{fig:loops}(a). 
The emergence of a common shape upon the scaling of the images suggests the existence of a mathematical similarity solution of the loop profile. 

We notice that at high pressure, the loops become small with large curvature so that the total energy Eq.~\ref{eq:U_tot} of our model filament is dominated by the bending term $U\ts{b}$ and the pressure term $U\ts{p}$. 
Minimizing $U\ts{b}+U\ts{p}$ (neglecting the pinning potential $U\ts{t}$), we obtain a differential equation for the loop shape (see SI and~\cite{Landau86, Flaherty72, Py07a}). 
We solve this equation numerically and plot the result as a solid curve on Fig.~\ref{fig:loops}(a), which is in excellent agreement with the composite experimental image. 

Having settled the shape of the loops, we now investigate their size. 
Intuitively, we expect a balance between the bending and pressure effects to determine their width, leading to: 
\begin{align}
w\textsubscript{loop}=\xi\bigg(\frac{B}{p}\bigg)^{1/3} .
\label{eq:scaling}
\end{align}
This scaling is supported by the full analysis of the filament model (see SI), which further gives the value of the prefactor, $\xi=1.065$. 
%where $\xi=1.065$ is a dimensionless coefficient that we determine from analyzing the filament model numerically (see SI). 
We measure the width of many loops at two different cross-section heights, and at multiple values of air pressure.  
The blue symbols in Fig.~\ref{fig:loops}(b) show the results, where the error bars are computed as the standard deviation of the mean. 
The data at two different $z$-positions, represented by two different shades of blue, fall on top one another, once again supporting the local treatment of the cross-section model. 
All the data are in good agreement with the predicted scaling of Eq.~\ref{eq:scaling} with pressure. % ($w\textsubscript{loop} \propto p^{-1/3}$). 
%and the data compares well to the predicted $1/3$-power law
% of $l\textsubscript{loop}\sim p^{-0.37\pm0.02}$, as 
%shown in Fig.~\ref{fig:loops}(b).  

As a more stringent test, we may compare Eq.~\ref{eq:scaling} directly with the data, by using a value of $B=(1.58\pm0.06)\times 10^{-8}$ J for the balloon material, that we obtained by measuring the Young's modulus $E$ using a tensile tester, and using $B = E t^3/[12 (1-\nu^2)]$ with $\nu = 0.4$ (see SI). 
The prediction is in good agreement with the data [red line in Fig.~\ref{fig:loops}(b)]. 
Conversely, one can use a power-law fit to the data to measure the bending modulus of the film. 
For our data, this yields a value of $B=(1.44\pm0.15)\times 10^{-8}$ J, which is a surprisingly accurate measurement. %, without prior knowledge of the Poisson's ratio. 

Finally, we compare our results with the cross-section model, which incorporates the confinement term $U\ts{t}$ in Eq.~\ref{eq:U_tot}. 
We generate cross sections at multiple pressures and measure the width of the loops. 
The result [red markers in Fig.~\ref{fig:loops}(b)] agrees well with the experiments.

\section{Discussion}

We have probed the gross shape, the fine structure, and the interplay between them in a balloon made of two flat elastic disks sealed at the edge, partly filled with water and pressurized. 
We have considered the geometric model that treats the gross shape of the sheet as inextensible yet free to compress, due to a vanishing bending modulus that allows for the formation of buckled fine structures at negligible energetic cost. 
We found that this model predicts the ``mean shape'' of the sheet, defined as the mean radius of the cross sections. 
%In some cases, this mean shape can differ from the apparent shape. 
We have then shown that the undulations of the cross sections around the mean shape are well described by a phenomenological model that accounts for the resistance to bending, the effect of pressure, and a pinning to the mean shape. 
In particular, this model captures the transition between wrinkles resembling the Elastica at zero pressure to folds with a rounded tip when pressure is applied. 

A wrinkle-fold transition is usually associated with the global compression exceeding a finite threshold value~\cite{Pocivavsek08,Oshri15}. 
This threshold is predicted to be of the same scale of the wrinkle wavelength~\cite{Diamant11}, so that wrinkles typically become unstable to forming a fold at small slopes. 
Here, at zero pressure, we observe wrinkles that are stable up to large slopes, and at a global compression that is much larger than the wavelength. 
Notably, in our system the transition to folding occurs as the azimuthal compression \textit{decreases}, signaling a different mechanism than the well-studied wrinkle-fold transition that is driven by a growing wrinkle amplitude. 
Indeed, the transition to folds that we observe is due to the internal pressure that deforms the wrinkles, leading to self contact. 
The ability of wrinkles to survive at large amplitude is rooted in the restoring force when the cross section deviates from its ideal position, which has components both in the radial and angular directions. % Is this a spot to clarify/remind why there is an angular restoring force?

There is another known class of wrinkle-fold transitions with a distinct geometric mechanism. 
In some settings, the energetically-favored gross shape may break axisymmetry in such a way where some of the excess length must be stored locally. 
This scenario may be compatible with the formation of localized deep folds but not with regular wrinkles~\cite{Paulsen15,Paulsen17}. 
Our results show that a pressurized water balloon does not fall into this class: both wrinkles and folds serve to waste excess material length along an axisymmetric gross shape. 
Hence, our system displays a novel type of wrinkle-fold transition where wrinkles are stabilized by an isotropic tension term but destabilized by the internal pressure. 

More generally, our results raise fundamental questions about the nature of convergence towards the shape predicted by the geometric model. 
The geometric model predicts the limiting shape of the sheet as the bending modulus $B$ vanishes.
This limiting shape has a compressive strain, which corresponds to infinitesimal undulations of the actual sheet around its mean shape \cite{Tobasco21}. 
Previous works, which have focused on nearly planar shapes with small effective compressive strain, have suggested that these undulations usually take the form of sinusoidal wrinkles~\cite{Davidovitch11,King12}. 
Our observations show that, when the effective compressive strain is large, the undulations could also take the form of deep folds.
Beyond the shape of the undulations, the convergence towards the prediction of the geometric model can be quantified by the evolution of the wavelength $\lambda$ as $B\to 0$.
For the tensional wrinkles that we observe, the wavelength should scale as $\lambda\sim B^{1/4}$~\cite{Cerda03}.
For the tensional folds, a scaling argument leads to $\lambda_\text{fold} \sim B^{2/9}$ (SI) for the distance between folds. 
Since the loops at the tips of the folds follow $w_\text{loop} \sim B^{1/3}$, then $\lambda_\text{fold} \gg w_\text{loop}$ as the bending modulus vanishes, ensuring that the folds remain spatially separated in this limit.

\begin{acknowledgments} 
We thank Pan Dong for assistance with the tensile tester. 
We thank the Syracuse Biomaterials Institute for use of the tensile tester. 
This work was supported by NSF Grant No. DMR-CAREER-1654102 (M. H. and J. D. P.). 
\end{acknowledgments}

%\bibliography{bibsheets}

%\end{document}

%%%%%%%%%%%%%%%%%%%%%%%%%%%%%%%%%%%%%%%%%%
%%%%%%%%%%%%%%%%%%%%%%%%%%%%%%%%%%%%%%%%%%
%%%%%%%%%%%%%%%%%%%%%%%%%%%%%%%%%%%%%%%%%%
%%%%%%%%%%%%%%%%%%%%%%%%%%%%%%%%%%%%%%%%%%
%%%%%%%%%%%%%%%%%%%%%%%%%%%%%%%%%%%%%%%%%%
%%%%%%%%%%%%                                               %%%%%%%%%%%%%%%
%%%%%%%%%%%%                       SI                     %%%%%%%%%%%%%%%
%%%%%%%%%%%%                                               %%%%%%%%%%%%%%%
%%%%%%%%%%%%%%%%%%%%%%%%%%%%%%%%%%%%%%%%%%
%%%%%%%%%%%%%%%%%%%%%%%%%%%%%%%%%%%%%%%%%%
%%%%%%%%%%%%%%%%%%%%%%%%%%%%%%%%%%%%%%%%%%
%%%%%%%%%%%%%%%%%%%%%%%%%%%%%%%%%%%%%%%%%%
%%%%%%%%%%%%%%%%%%%%%%%%%%%%%%%%%%%%%%%%%%

%\newpage
\clearpage

\onecolumngrid

\renewcommand{\thefigure}{S\arabic{figure}}
\setcounter{figure}{0} 
\renewcommand{\theequation}{S\arabic{equation}}
\setcounter{equation}{0}
\renewcommand{\thesection}{SI \arabic{section}}
\setcounter{section}{0} 

\noindent\textbf{\LARGE Supplementary Information for \\} \\
%\noindent\textbf{Supplementary Information for \\
\noindent\textbf{\Large ``Interplay of gross and fine structures in strongly-curved sheets" \\} \\
%\end{center}
\noindent\textbf{\footnotesize Mengfei He, Vincent D\'emery, Joseph D. Paulsen}
\vspace{0.75in}

\section{Experimental methods}

\begin{figure}[htp]
	\includegraphics[width=0.9\textwidth]{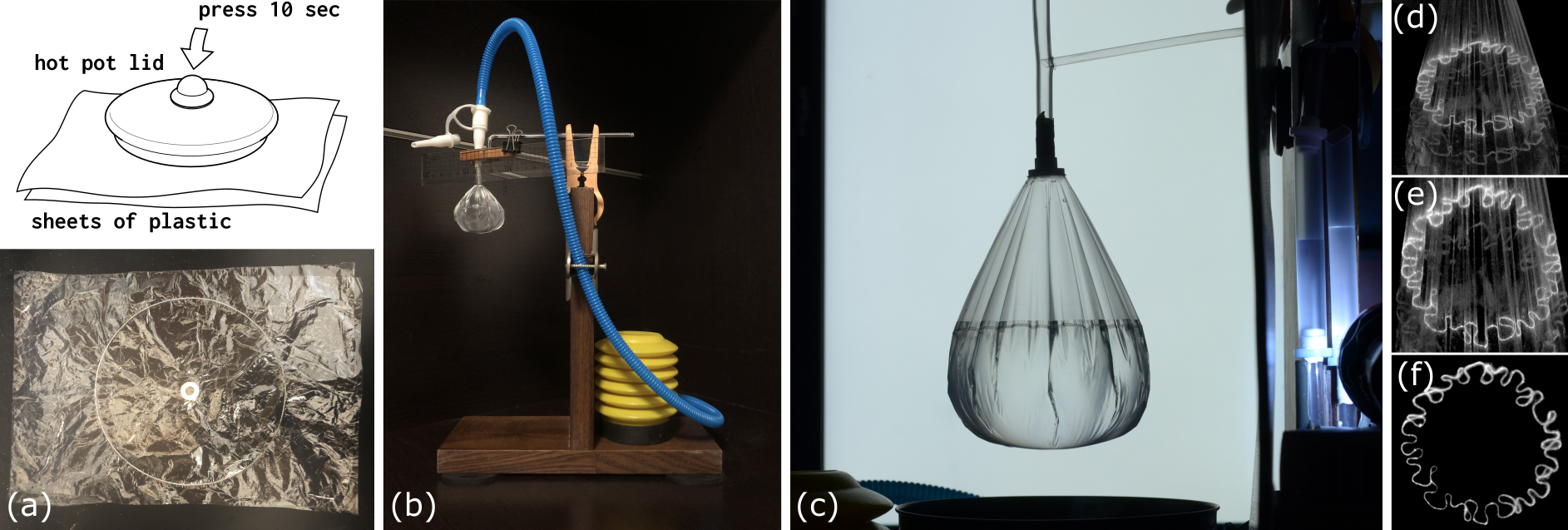}
	\centering
	\caption{Experimental setup for a water balloon system.  (a) A heated circular metal hoop is pressed against two aligned plastic sheets, locally heat fusing the sheets at the circular edge. The fused double layers are then cut out to form the initial state of a balloon.  (b) The balloon is partially filled with water, pumped with air and hung from the top.  (c) An LED monitor provides uniform back-light for profile photography.  A U shaped tube is connected with the balloon with the water column difference indicating the internal air pressure of the balloon.  (d) A laser sheet is scattered into the balloon to obtain a cross section, viewed from an oblique angle.  A perspective transform brings the cross section into (e) The top view.  (f) Rotating and averaging results for the cross section removes noise.}
	\label{fig:setup}
\end{figure}

We use polyethylene produce bags or plastic food wrap as the balloon membrane. 
To construct the bags, two flat sheets are aligned at a flat rigid surface with a soft cushion of several layers of paper. 
To form an air tight seal, we press a heated iron pot lid with a sharp, circular edge ($\diameter=166$ mm), which forms a thin seam where the two layers are locally melted and bonded together. 
To homogenize the heating, parchment paper is inserted as a mask before pressing. 
By optimizing the heating and pressing time, we can obtain a seam width of $0.5\sim1$ mm. 
The fused circular double layers are then cut from the rest of the material as the initial flat state of the balloon. 
We locally treat the surface of the plastic films with a chemical primer to decrease the surface energy before gluing to a nylon washer at the top center [Fig.~\ref{fig:setup}(a)]. 
The system is then partially filled with water and suspended from the top (Fig.~\ref{fig:setup}(b)). 
We use an LED monitor to back-light the balloon uniformly, and U-shaped tube with water columns is connected with the inflated balloon to indicate the air pressure inside [Fig.~\ref{fig:setup}(c)]. 
To obtain the cross sections of the balloon, we use a laser sheet to scatter light horizontally into the system, which we record at an oblique angle to enhance visibility [Fig.~\ref{fig:setup}(d)]. 
We use a calibrated perspective transformation (OpenCV-Python) to bring the cross sections into a top view [Fig.~\ref{fig:setup}(e)]. 
We superimpose thus obtained cross sections from different angles by azimuthually rotating the balloon. 
The high overlap of the signals in the superimposed image corroborates the accuracy of the perspective transformation, while the averaging removes almost all noise from secondary scatterings [Fig.~\ref{fig:setup}(f)].

\begin{figure}[htp]
	\includegraphics[width=0.95\textwidth]{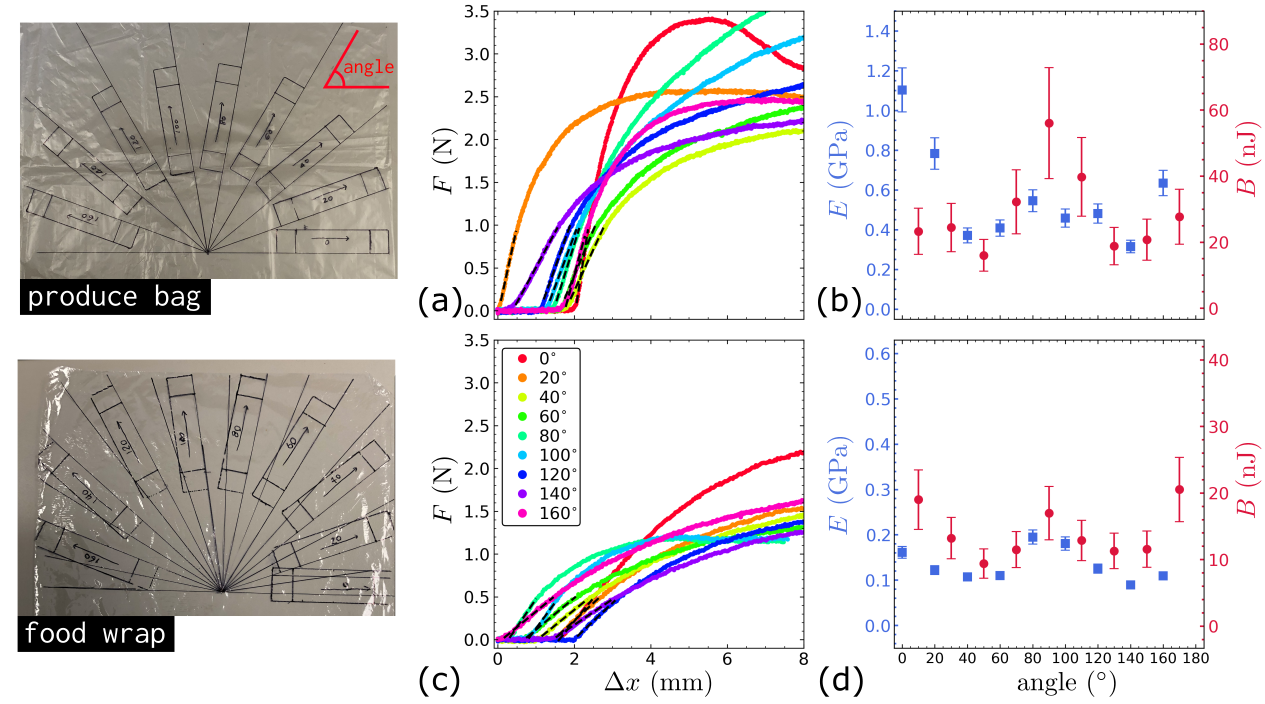}
	\centering
	\caption{Images: Rectangular samples are cut from the sheets at various orientations and loaded into a tensile tester for modulus measurements.  (a, c) Respective force-displacement curves for samples cut from a produce bag and a food wrap used in the experiments.  Dashed lines: Linear fits applied at $0.1$ N $<F<1$ N (produce bag) and $0.03$ N $<F<0.5$ N (food wrap) to compute (b, d) Young's modulus $E$ and the bending modulus $B$ for the samples, as functions of the orientation.  Error bars propagated from the uncertainty in the thickness measurements.}
	\label{fig:EB}
\end{figure}

We characterize the plastic sheets used in our experiments by measuring their thickness and Young's modulus. 
The thickness $t$ of the balloon membranes is measured by using a caliper (Mitutoyo) for multiple layers with a linear fit to be $8\pm0.8$ $\mu$m (produce bag) and $10\pm0.8$ $\mu$m (food wrap). 
The plastic films have an anisotropic modulus due to the manufacturing process. 
Due to system geometry, in the experiment, wrinkles and lines of tension emanate from the top and bottom of the balloon radially.
As a result he elastic moduli of a sheet along all directions become relevant, which we measure with a tensile tester (Test Resources, 250 lbs actuator). 
To characterize the anisotropy, we cut rectangular samples of $16.5$ mm $\times50$ mm from the sheets at various orientations (Fig.~\ref{fig:EB}, images).
Each sample is characterized measuring its stress-displacement curve. 
Linear fits are carried out in the regions of $0.1$ N $<F<1$ N (produce bag) and $0.03$ N $<F<0.5$ N (food wrap), as shown by the dashed lines in Fig.~\ref{fig:EB}(a,c).  The corresponding Young's modulus $E$ and the bending modulus $B\equiv Et^3/[12(1-\nu^2)]$ are calculated from the fits, shown in Fig.~\ref{fig:EB}(b), (d).  We use a Poisson's ratio of $\nu=0.4$ for both materials. 
The errors in $B$ are dominated by the uncertainty in the thickness $t$, which we propagate to the error bars in Fig.~\ref{fig:EB}(b,d).

\section{Model for the cross section of a water balloon}
\subsection{Energy}\label{}

We model the cross section as an inextensible filament with length $L\equiv2\pi \ls$ described by $\rr(s)$.
We write the energy as
\begin{equation}\label{eq:energy}
U = \frac{B}{2}\int \rr''(s)^2\dd s + \frac{T}{2}\int [\rr(s)-\rr\ind{id}(s)]^2\dd s - \frac{p}{2}\int \rr(s)\times\rr'(s)\dd s,
\end{equation}
where $B$ is the bending modulus, $T\equiv\eta T\ind{s}/r\ind{disk}^2$ the tension, $p$ the pressure, and $\rr\ind{id}(s)$ is the ``ideal'' position of the cross section, which is a circle with radius $r\ind{conf}$.
Here the tension plays the same role as in~\cite{Cerda03_SI}, and is not the tension in the filament.  The dimensionless scaling factor $\eta$ in the tension term is to be determined by matching the wavenumber produced by the model with the experimental measurement, as shall be discussed later.

The force in the filament is given by
\begin{equation}
\ff(s) = -\frac{\delta U}{\delta \rr(s)} = -B\rr^{(4)}(s)+T[\rr\ind{id}(s)-\rr(s)]-p\rr_\perp'(s).
\end{equation}

Our goal is to minimize the energy by writing a dynamics of the form
\begin{equation}\label{eq:dynamics}
\dot\rr(s) = \ff(s).
\end{equation}
Before applying this dynamics, we have to include repulsion between distant parts of the filament and add a Lagrange multiplier to satisfy inextensibility.

\subsection{Geometric radius}\label{}

If the bending modulus goes to 0, the cross section makes tiny undulations around a circle with radius $R$.
In this case, the energy (\ref{eq:energy}) of a cross section reduces to
\begin{equation}
U = \pi T (R-r\ind{conf})^2\ls-\pi pR^2.
\end{equation}
Minimizing the energy leads the radius
\begin{equation}
R^* = \frac{r\ind{conf}}{1-p/(T\ls)}.
\end{equation}
This corresponds to the radius predicted by the geometric model, $R^*=r\ind{gm}$.
As a consequence, in the presence of pressure, the confinement radius to be used in the model is given by $r\ind{conf}=[1-p/(T\ls)]r\ind{gm}$.

\subsection{Implementation in the discrete model}\label{}

We discretize the filament with $N$ nodes with position $\rr_i$, $i\in \{0,\dots,N-1\}$.
Nodes are linked by segments with rest length $a=L/N$.
%, where $L$ is the rest length of the filament.
%Below we give the different terms entering the force $\ff_i$ that applies on each node:
Below we give the different terms entering the force $\ff_i$ that applies on each node:
The force $\ff_i$ that applies on each node contains contributions from bending, tension, pressure, repulsion and inextensibility:
\begin{equation}
\label{eq:terms}
\ff = \ff\ind{b}+\ff\ind{t}+\ff\ind{p}+\ff\ind{r}+\ff\ind{i}.
\end{equation}
%where different terms represent the bending, tension, pressure, repulsion and inextensibility forces.

\subsubsection{Bending}\label{}

To determine the bending force in the discrete filament, we define a bending energy that depends on the angle between neighboring segments.
Denoting $\theta_i$ the angle between $\rr_{i+1}-\rr_i$ and $\rr_i-\rr_{i-1}$, the bending energy is
\begin{equation}
U\ind{b} = A\sum_i f(\theta_i),
\end{equation}
where $f(\theta)$ is quadratic for small angles, $f(\theta)\sim \theta^2/2$.

This energy should correspond to the continuous version $U\ind{b}=(1/2)\int\rr''^2$.
To determine the value of the constant $A$, we equate the continuous and discrete energies for a filament that describes a circle with radius $R$.
The continuous bending energy is $U\ind{b}=\pi B/R$, while the discrete energy is
\begin{equation}
U\ind{b}=AN f \left(\frac{2\pi}{N} \right)\simeq \frac{2\pi^2 A}{N},
\end{equation}
Equating the two expressions, we get
\begin{equation}
A=\frac{B}{a}.
\end{equation}

From the positions of the nodes, the angles are calculated by
\begin{equation}
\cos(\theta_i)=\hat\ee_i\cdot \hat\ee_{i-1},
\end{equation}
where we have defined the unit vectors
\begin{equation}
\hat\ee_i = \frac{\rr_{i+1}-\rr_i}{|\rr_{i+1}-\rr_i|},
\end{equation}
where $|\rr|$ is the norm of the vector $\rr$.
Instead of inverting the cosine function, we write $f(\theta)$ as
\begin{equation}
f(\theta)=g(\cos(\theta)).
\end{equation}
The condition $f(\theta)\sim\theta^2/2$ for small $\theta$ translates into 
\begin{equation}\label{eq:constraint_gbend}
g'(1)=-1;
\end{equation}
the constraint on $f(\theta)$ also implies $g(1)=0$, but this condition does not have to be satisfied since the energy is defined up to a constant.

The bending force is given by the variation of the energy with respect to the position of the nodes.
This calculation requires to know the variation of $\hat\rr=\rr/|\rr|$ with $\rr$, which is
\begin{equation}
\frac{\delta\hat\rr}{\delta\rr}= \frac{\hat\rr_\perp\otimes\hat\rr_\perp}{|\rr|},
\end{equation}
where the subscript $\perp$ denotes a rotation by an angle $\pi/2$.
The bending force on the node $i$ is thus
\begin{align}
\ff_{\mathrm{b},i} &= -A \left[g'(\cos(\theta_{i-1})) \frac{\delta\cos(\theta_{i-1})}{\delta\rr_i} 
+ g'(\cos(\theta_{i} )) \frac{\delta\cos(\theta_{i})}{\delta\rr_i} 
+ g'(\cos(\theta_{i+1})) \frac{\delta\cos(\theta_{i+1})}{\delta\rr_i}\right]\\
& = -A \left[g'(\cos(\theta_{i-1})) \frac{\hat\ee_{i-1,\perp}\cdot\hat\ee_{i-2}}{|\rr_i-\rr_{i-1}|}\hat\ee_{i-1,\perp} 
- g'(\cos(\theta_{i+1})) \frac{\hat\ee_{i,\perp}\cdot\hat\ee_{i+1}}{|\rr_{i+1}-\rr_{i}|}\hat\ee_{i,\perp} \right.\nonumber
\\&\qquad\qquad\left. + g'(\cos(\theta_{i} )) \left(\frac{\hat\ee_{i-1,\perp}\cdot\hat\ee_{i}}{|\rr_i-\rr_{i-1}|}\hat\ee_{i-1,\perp} - \frac{\hat\ee_{i,\perp}\cdot\hat\ee_{i-1}}{|\rr_{i+1}-\rr_{i}|}\hat\ee_{i,\perp} \right) \right]
\end{align}

The simplest function $g(x)$ satisfying the constraint (\ref{eq:constraint_gbend}) is $g(x)=-x$. 
However, this choice does not penalize large angles between neighboring segments, which we have to avoid; in particular, the torque at a node vanishes as the angle approaches $\pi$.
We choose instead
\begin{equation}
g(x)=\frac{4}{1+x},
\end{equation}
which diverges as an angle approaches $\pi$. 
Its derivative is $g'(x)=-4/(1+x)^2$.

\subsubsection{Tension}\label{}

We discretize the tension energy as:
\begin{equation}
U\ind{t} = \frac{aT}{2}\sum_i \left(\rr_i-\rr_{\mathrm{id},i} \right)^2,
\label{eq:Ut}
\end{equation}
where
\begin{equation}
\rr_{\mathrm{id},i} = r\ind{conf} \begin{pmatrix}
\cos(2\pi i/N)\\ \sin(2\pi i/N)
\end{pmatrix}.
\end{equation}
The tension force is thus
\begin{equation}
\ff_{\mathrm{t},i} = aT\left(\rr_{\mathrm{id},i} -\rr_i\right).
\end{equation}

\subsubsection{Pressure}\label{}

The pressure energy is $U\ind{p}=-p\cA$, where $\cA$ is the area enclosed by the filament.
It is given by
\begin{equation}
\cA = \frac{1}{2}\sum_i \rr_i\times\rr_{i+1}= \frac{1}{2}\sum_i\rr_{i,\perp}\cdot\rr_{i+1}
=-\frac{1}{2}\sum_i\rr_{i}\cdot\rr_{i+1,\perp},
\end{equation}
assuming that the filament is oriented counter-clockwise.
As a consequence,
\begin{equation}
\frac{\delta\cA}{\delta\rr_i} = \frac{1}{2}(-\rr_{i+1,\perp}+\rr_{i-1,\perp}),
\end{equation}
hence
\begin{equation}
\ff_{\mathrm{p},i} = -\frac{p}{2}(\rr_{i+1}-\rr_{i-1})_\perp.
\end{equation}

\subsubsection{Repulsion}\label{}

Many choices are possible to avoid self-crossing while allowing two parts of the filament to slide against each other, see for instance Ref.~\cite{Evans13_SI}.

We do not derive the repulsion force from an energy.
Instead, for a given node $i$ we find the closest node $n(i)$, discarding the nodes $i\pm 1$.
For the segment $[n(i),n(i)+\nu]$, $\nu\in\{-1,1\}$, we find the point $\rr_\nu$ on this segment that is the closest to the node $i$.
This can be done by calculating
\begin{equation}
\lambda_\nu = \max\left(0,\min\left(1, \frac{(\rr_i-\rr_{n(i)})\cdot(\rr_{n(i)+\nu}-\rr_{n(i)})}{(\rr_{n(i)+\nu}-\rr_{n(i)})^2}\right)\right),
\end{equation}
leading to
\begin{equation}
\rr_\nu = \rr_{n(i)}+\lambda_\nu\left(\rr_{n(i)+\nu}-\rr_{n(i)}\right).
\end{equation}
We then choose the value of $\nu\in\{-1,1\}$ that minimizes the distance $|\rr_\nu-\rr_i|$, the point $\rr_\nu$ corresponding to this value is denoted $\rr_i^*$.

We compute a repulsion force on $\rr_i$ as
\begin{equation}
\pphi_i = \frac{\rr_i-\rr_i^*}{|\rr_i-\rr_i^*|}\Phi(|\rr_i-\rr_i|^*),
\end{equation}
with 
\begin{equation}
\Phi(r) = \theta(b-r) \left[\left(\frac{b}{r} \right)^7- \left(\frac{b}{r} \right)^4\right],
\end{equation}
where $\theta(u)$ is the Heaviside function and $b$ is the interaction range, which we take to be a fraction of $a$.

Finally, we remove the component of $\phi_i$ that is tangent to the filament at the node $i$, leading to the repulsion force
\begin{equation}
\ff_{\mathrm{r},i}= \pphi_i-\frac{(\rr_{i+1}-\rr_{i-1})\cdot\pphi_i}{(\rr_{i+1}-\rr_{i-1})^2} (\rr_{i+1}-\rr_{i-1}).
\end{equation}

%Following Ref.~\cite{Evans2013}, the normal force is reduced and inverted when two parts of the filament come close to each other.
%We obtain a corrected force $\ff(s)$.

\subsubsection{Inextensibility}\label{}

The dynamics (\ref{eq:dynamics}) has to satisfy inextensibility, which can be written as $\rr'(s)\cdot\ff'(s)=0$; this is ensured by the inextensibility force $\ff\ts{i}$.
We denote the sum of all the other terms of the force as $\ggg= \ff\ind{b}+\ff\ind{t}+\ff\ind{p}+\ff\ind{r}$.

%The force $\ff_i$ that applies on each node contains contributions from bending, tension, pressure, repulsion and inextensibility:
%\begin{equation}
%\label{eq:terms}
%\ff = \ff\ind{b}+\ff\ind{t}+\ff\ind{p}+\ff\ind{r}+\ff\ind{i}.
%\end{equation}

To define the ``inextensibility force'' $\ff\ts{i}$, we introduce the tension in the filament $\tau(s)$, so that $\ff\ts{i}$ is the force due to this tension~\cite{Tornberg04_SI}:
\begin{equation}
\ff\ind{i}(s) = (\tau\rr')'(s).
\end{equation}
%The equation to be satisfied by the tension is thus
%\begin{equation}
%\rr'\cdot(\tau\rr')'=-\rr'\cdot\ggg'.
%\end{equation}
%This is a second order ODE.
Actually, as strict inextensibility is difficult to ensure, we instead impose that
\begin{equation}
\rr'\cdot\ff'=Y\left(1-{\rr'}^2\right),
\end{equation}
where $Y$ is a large stretching modulus; with this term, any deviation to inextensibility quickly relaxes to zero.
The equation for the tension is thus
\begin{equation}
\rr'\cdot(\tau\rr')'=-\rr'\cdot\ggg'+Y\left(1-{\rr'}^2\right).
\end{equation}
Instead of discretizing this equation, we write directly the equation for the discretized variables.

We denote $\rr_i$ the position on the node $i$ and $\ff_i$ the force on this node.
The evolution of the length of the segment $[i,i+1]$, which should be equal to $Y(1-\rr'^2)$, is given by 
\begin{equation}\label{eq:evol_length_disc}
\rr'\cdot\ff' = a^{-2}(\rr_{i+1}-\rr_{i})\cdot(\ff_{i+1}-\ff_{i}) = Y \left[1-a^{-2}(\rr_{i+1}-\rr_{i})^2 \right].
\end{equation}
%We write the total force as the initial force $\ggg=\ff\ind{b}+\ff\ind{t}+\ff\ind{p}+\ff\ind{r}$ and the tension force $\ff\ind{i}$: $\ff=\ggg+\ff\ind{i}$.
Denoting $\tau_i$ is the tension in the segment between the nodes $i$ and $i+1$, the tension force on the node $i$ is given by
\begin{equation}\label{eq:tension_force_disc}
\ff_{\mathrm{i},i} = a^{-2} \left[\tau_i(\rr_{i+1}-\rr_i)- \tau_{i-1}(\rr_{i}-\rr_{i-1})\right].
\end{equation}
Inserting this expression in Eq.~(\ref{eq:evol_length_disc}) leads an equation for the tension:
\begin{multline}
(\rr_{i+1}-\rr_i)\cdot(\rr_{i+2}-\rr_{i+1}) \tau_{i+1}-2(\rr_{i+1}-\rr_i)^2 \tau_i+(\rr_{i+1}-\rr_i)\cdot(\rr_{i}-\rr_{i-1}) \tau_{i-1} \\
= Y a^2 \left[a^2-(\rr_{i+1}-\rr_i)^2 \right]-  a^2(\rr_{i+1}-\rr_i)\cdot(\ggg_{i+1}-\ggg_i).
\end{multline}

Last, we define $\tau_i=a^2\tilde \tau_i$ and put the equation above under the form
%Defining $\tau_i=a^2\tilde \tau_i$, 
%the tension force is
%\begin{equation}\label{eq:tension_force_disc2}
%\ff_{\mathrm{i},i} = \tilde \tau_i(\rr_{i+1}-\rr_i)- \tilde \tau_{i-1}(\rr_{i}-\rr_{i-1}),
%\end{equation}
%and it is the solution of
\begin{equation}\label{eq:tension_matrix}
\sum_{j} A_{ij}\tilde \tau_j = B_i,
\end{equation}
with
\begin{align}
A_{i,i+1} & = (\rr_{i+1}-\rr_i)\cdot(\rr_{i+2}-\rr_{i+1}) ,\\
A_{i,i} & = -2(\rr_{i+1}-\rr_i)^2,\\
A_{i,i-1} & = (\rr_{i+1}-\rr_i)\cdot(\rr_{i}-\rr_{i-1}),\\
B_i & = Y \left[a^2-(\rr_{i+1}-\rr_i)^2 \right]-(\rr_{i+1}-\rr_i)\cdot(\ggg_{i+1}-\ggg_i).
\end{align}
Solving equation (\ref{eq:tension_matrix}) gives the tension and thus the tension force $\ff\ts{i}$. 
This conclude our definition of the filament dynamics.

\begin{figure}[htp]
	\includegraphics[width=0.95\textwidth]{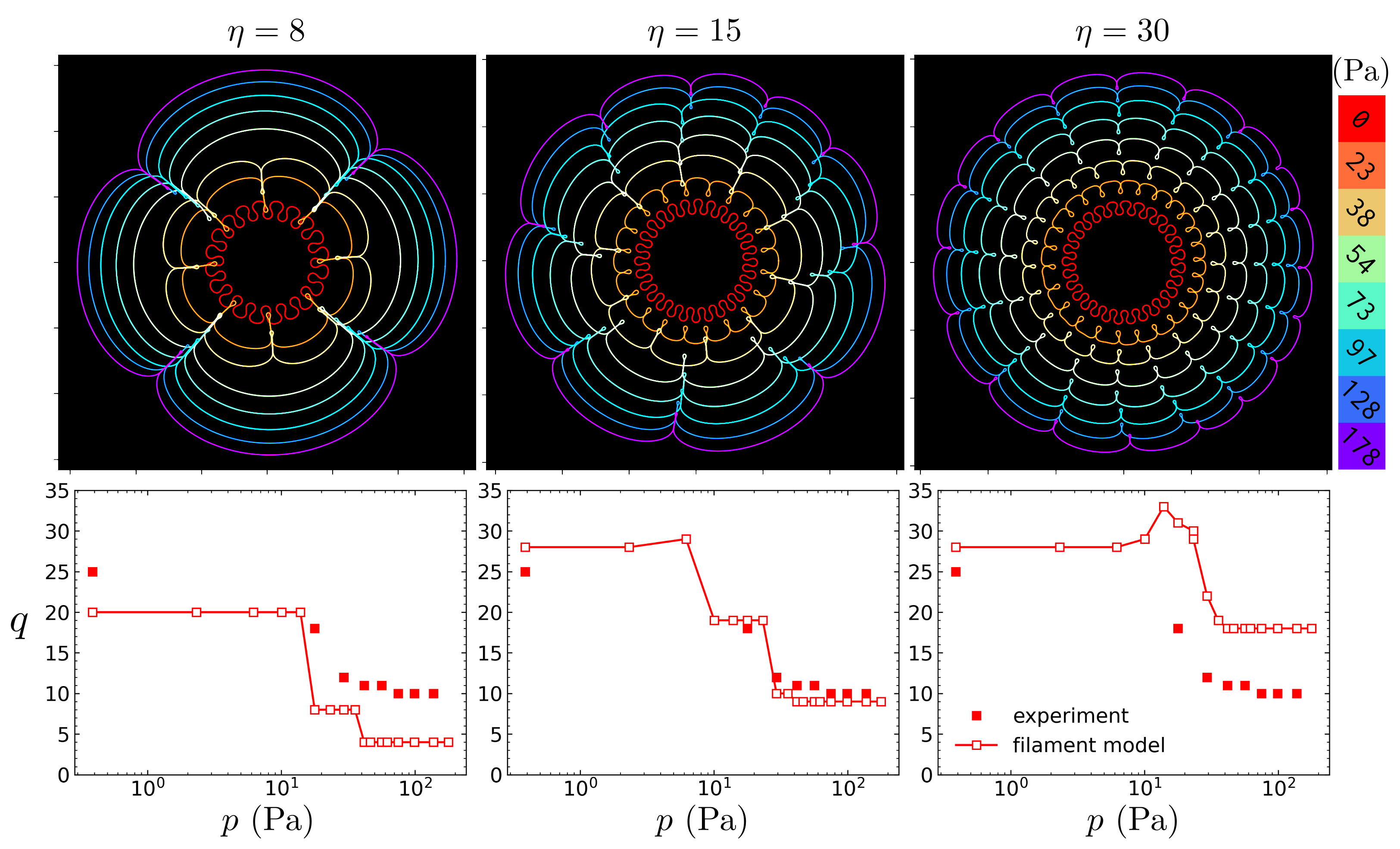}
	\centering
	\caption{Cross-section configurations given by the filament model with different selections of the $\eta$ coefficient in Eq.~\ref{eq:energy}.  Increasing $\eta$ increases the overall wavenumber $q$, and shifts the $q(p)$ curve vertically with the wrinkle-fold transition taking place in the same $p$ region.  We choose $\eta=15$ to best match the wavenumber with the experiments at all pressures (middle panel).}
	\label{fig:eta}
\end{figure}

\subsection{Equilibrium configurations and the selection of $\eta$}
Finally, various terms of forces in Eq.~\ref{eq:terms} are collected and the dynamics Eq.~\ref{eq:dynamics} is evolved using the forward Euler method with a step size such that $\max(\abs{\dot\rr_i})\Delta t\le 0.05a$.
We use a small interaction range $b=0.2a$ for the node-segment repulsion, and a large numerical value $10^6$ for the effective stretching modulus $Y$.  We start with a high pressure for $N=1000$ nodes and the initial configuration of $\rr_i=l\ts{s}[\cos(2\pi i/N), \sin(2\pi i /N)]$ with a fluctuation of $0.1a$.

Figure~\ref{fig:eta} shows the evolution of the modeled cross section of a deflating balloon from the air pressure $p=178$ Pa to $p=0$, for three different fixed $\eta$ values.  Regardless of the choice of $\eta$, the model qualitatively captures the lobed shape at a high pressure with deep folds,  a rounded wrinkled shape when fully deflated, and a wrinkle-fold transition with an increase of the wavenumber at an intermediate pressure.  An increase of $\eta$ mainly shifts the $q(p)$ curve upwards, while having a minimal effect on the pressure region of the wrinkle-fold transition.  We therefore conclude that our system is insensitive to $\eta$ in terms of the wrinkle-fold transition, and we choose the value of $\eta=15$ which best matches with the experimentally measured wavenumber of the system at all pressures.

\section{Asymptotic analysis of loops at high pressure}

At high pressure, deep folds in between the lobes form with loops at the tips.  The loops are small in size so that the bending and pressure terms in Eq.~\ref{eq:terms} dominate.  Below we analyze such asymptotic regime of a small two-dimensional loop where there is a simplified balance between bending and pressure.

\begin{figure}[htp]
	\includegraphics[width=0.9\textwidth]{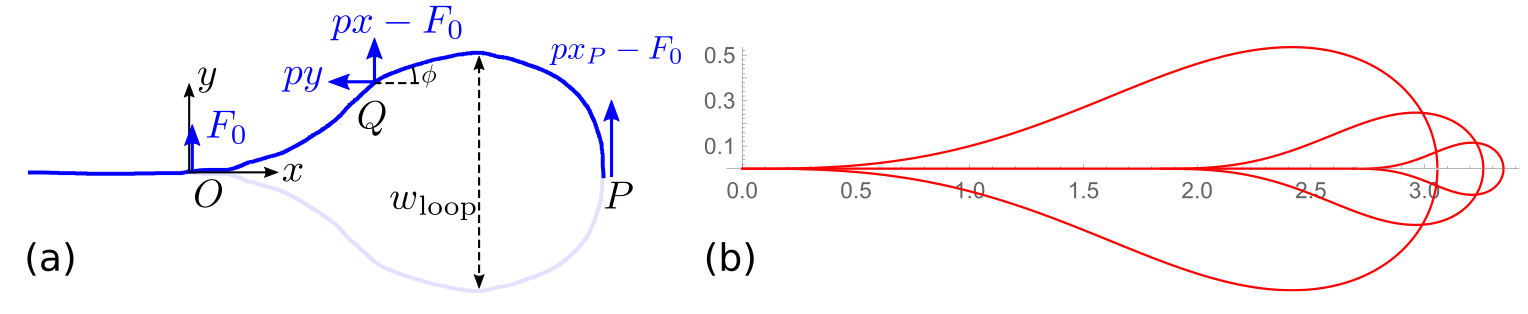}
	\caption{
	(a) Schematic of a sheet forming a loop at the tip of a fold, under a high external pressure, where the tension term $U\textsubscript{t}$ in Eq.~\ref{eq:energy} can be neglected.
	(b) Numerical solutions for an asysmptotic loop at $p/B=1,10,100$ (with the consistent unit of length$^{3}$).}
	\label{fig:loop}
\end{figure}

Focusing on the upper part of a loop [Fig.~\ref{fig:loop}(a)], the internal force at the tip $P$ must be vertical due to symmetry.  As a result, the force $F_0$ at the contact point $O$ must be vertical due to the horizontal force balance for $\wideparen{OP}$. We choose the origin of the coordinate system to be at this contact point $O$.  For a segment $\wideparen{OQ}$ along the arc, the force exerted by external pressure $p$ is $(yp,-xp)$.  Therefore, force balance for $\wideparen{OQ}$ gives the internal stress at $Q$ to be $(-yp, xp-F_0)$, as shown in the schematic.

Applying torque balance $d\textbf{M}/dl = \textbf{F}\times \textbf{t}$~\cite{Landau86_SI} at point $Q$ with the tangent $\bf{t} = (\cos\phi, \sin\phi)$ and $\textbf{M}= B(d\phi/dl)\bf{\hat{z}}$ results in the governing equation for the loop shape:
\begin{align}
	B\cos^2\phi\frac{d^2\phi}{dx^2}-B\cos\phi\sin\phi\bigg(\frac{d\phi}{dx}\bigg)^2= -yp\sin\phi -(xp-F_0)\cos\phi.
	\label{eq:torque_balance}
\end{align}

The contact force at point $O$ to the upper part of the loop has to be pointing upwards, so that $F_0$ must be positive.  Applying a rescaling $\textbf{x}\rightarrow \textbf{x}(F_0/p)$ and substituting $dy/dx = \tan \phi$ leads to a dimensionless ODE, to be solved numerically:
\begin{align}
	\phi'''-(3\tan\phi+\csc\phi\sec\phi)\phi'\phi''+\tan^2\phi\phi'^3+c(1-x)\csc\phi\sec^2\phi\phi'+c\sec^3\phi=0,
	\label{eq:3rd_Fnegative}
\end{align}
where 
\begin{align}
c \equiv \frac{F_0^3}{p^2B}
\label{eq:c}
\end{align}
is a dimensionless parameter.
%\begin{align}
%	&\textbf{x}\rightarrow \textbf{x}(-F_0/p).\label{eq:scaling}\\
%	&\frac{\cos^2\phi}{\sin\phi} \phi''-\cos\phi\phi'^2+c\frac{\cos\phi}{\sin\phi}(1-x)= cy
%	\label{eq:2nd_Fnegative}
%\end{align}

\subsection{Boundary conditions}
We seek a loop solution with a long, flat ``neck'' as the contact region extending to $x<0$.  First, $\phi(0)=0$ since otherwise there is infinite curvature at the contact line requiring infinite bending moment. However, notice Eq.~\ref{eq:3rd_Fnegative} is singular at $\phi=0$ so instead we set 
\begin{align}
\phi(0)=\epsilon,
\label{eq:bc0}
\end{align}
expecting the solution $\phi(x)$ to be independent of $\epsilon$ as $\epsilon$ approaches 0.  Next, integrating Eq.~\ref{eq:3rd_Fnegative} once gives
\begin{align}
	\label{eq:2nd_Fnegative}
	\phi'' &= -cy\frac{\sin\phi}{\cos^2\phi}+\phi'^2\frac{\sin\phi}{\cos\phi}-c\frac{(x-1)}{\cos\phi}\\&\xrightarrow{\phi=\epsilon}c.
	\label{eq:bc2}
\end{align}
The third condition is given by $y(0)=0$. Equations~\ref{eq:2nd_Fnegative} and~\ref{eq:bc2} give $cy(0)=\cos\epsilon[c\tan\frac{\epsilon}{2}+\phi'^2(0)]$, which leads to
\begin{align}
	\phi'(0) = \sqrt{-c\tan\frac{\epsilon}{2}} ,
	\label{eq:bc1}
\end{align}
so that $\phi'(0) \rightarrow 0$ as $\epsilon \rightarrow 0^{-}$ (the film curvature is zero at the point $O$).

\subsection{Similarity solution}\label{sec:similarity}
Admitting a similarity solution, the dimensionless Eq.~\ref{eq:3rd_Fnegative} is solved numerically using the boundary conditions Eqs.~\ref{eq:bc0}~\ref{eq:bc2}~\ref{eq:bc1}.  The only remaining parameter, $c$ is identified by requiring $y(x_p)=0$ where $x_p$ is the $x$-position of the loop tip and is given by $\phi(x_p)=-\pi/2$, which results in
\begin{align}
	c=0.587824\dots,
	\label{eq:c_value}
\end{align}
as a dimensionless geometric constant, independent of mechanical properties of $p$, $B$, $F_0$.  The resulting similarity solution with a "neck" is plotted in Fig.~\ref{fig:loop}(b) in the dimensional form (with arb. unit of length) with $p/B$ varying from 1 to 100 (with the consistent unit of length$^3$), sharing the same total length including the neck.  The similarity solution numerically gives the loop size [Fig.~\ref{fig:loop}(a)] $w\ts{loop}= 1.271(F_0/p)$.  Substitute in Eqs~\ref{eq:c}~\ref{eq:c_value}, we get
\begin{align}
	w\ts{loop} =1.065\bigg(\frac{B}{p}\bigg)^{1/3}.	
	\label{eq:wloop}
\end{align}
%This scaling behavior can also be seen from an energy argument: the balance between the bending energy $\sim\delta(\pi w\ts{loop}B/w\ts{loop}^2)$ and the virtual work done by the pressure $\sim\pi w\ts{loop} p \delta w\ts{loop}$ gives $w\ts{loop}^3\sim B/p$.
This scaling behavior can also be seen from an energy argument: the balance between the bending energy $B/w\ts{loop}$ and the work done by the pressure $p w\ts{loop}^2$ gives $w\ts{loop}\sim (B/p)^{1/3}$.

Similarly, for the contact force
\begin{align}
	F_0 = c^{1/3}B^{1/3}p^{2/3}.
	\label{eq:contact_force}
\end{align}
Finally, the energy of a loop scales as
\begin{equation}\label{eq:energy_loop}
u\ts{loop} \sim B^{2/3}p^{1/3}.
\end{equation}

\subsection{Distance between folds}\label{}

%\subsection{Without pressure}\label{}
%\begin{itemize}
%\item Bending energy per unit length: $B/\lambda^2$. 
%\item Tension energy per unit length: $T\lambda^2$, assuming the effective strain $\epsilon\sim 1$, hence the amplitude $A\sim \lambda$. 
%\end{itemize}
%Balancing the two leads to the Cerda and Mahadevan result~\cite{Cerda03}:
%\begin{equation}
%\lambda\sim \left(\frac{B}{T} \right)^{1/4}.
%\end{equation}
%\subsection{With pressure}\label{}
%We assume that the morphology here consists in lobes of width $\lambda$ separated by deep folds terminated by loops of width $\ell$.
%The size of the loops are obtained by balancing the bending energy, $\ell\times (B/\ell^2)\sim B/\ell$, with the pressure energy, $p\ell^2$, leading to 
%\begin{align}
%\ell & \sim (B/p)^{1/3},\\
%u\ind{loop} & \sim p^{1/3} B^{2/3},
%\end{align}
%where $u\ind{loop}$ is the energy of one loop, hence the energy of one fold.

We evaluate the distance between folds, $\lambda\ts{fold}$.
With an effective strain $\epsilon$, the angular and radial displacement in a lobe separating two folds is typically $\epsilon\lambda\ts{fold}$, leading to a tension energy per lobe $u\ts{lobe}\sim\lambda\ts{fold}\times T(\epsilon\lambda\ts{fold})^2\sim T\epsilon^2\lambda\ts{fold}^3$.

For a cross section of length $L$, there are $n=L/\lambda\ts{fold}$ lobes and folds, so that the total energy is
\begin{equation}
u=n(u\ts{lobe}+u\ts{loop})\sim L \left(T\epsilon^2\lambda\ts{fold}^2+ \frac{B^{2/3}p^{1/3} }{\lambda\ts{fold}} \right).
\end{equation}
Minimizing over $\lambda\ts{fold}$ leads to
\begin{equation}
\lambda\ts{fold}\sim \frac{p^{1/9}B^{2/9}}{T^{1/3}\epsilon^{2/3}}.
\end{equation}

\end{document}